\documentclass[amsmath,amssymb,floats,preprint,showpacs,preprintnumbers]{revtex4}
\usepackage[T1]{fontenc}
\usepackage[applemac]{inputenc} 	
\usepackage{graphicx}

\usepackage{dcolumn}
\usepackage{bm}

\def\be{\begin{equation}}
\def\ee{\end{equation}}
\def\bea{\begin{eqnarray}}
\def\eea{\end{eqnarray}}

\newcommand{\YBCO}{YBa$_2$Cu$_3$O$_{x}$}

\newcommand{\LSCO}{La$_{2-x}$Sr$_x$CuO$_{4}$}

\begin{document}
\title{Double criticality in the magnetic field-driven transition of a high-T$_C$ superconductor}
\author{Brigitte Leridon$^{1*}$, J. Vanacken$^{2}$,V.V. Moshchalkov$^{2}$, Baptiste Vignolle$^{3}$,  Rajni Porwal$^{4}$ and R.C. Budhani$^{4}$}

\affiliation{$^1$Laboratoire de Physique et  d'Etude des Mat\'eriaux,  UMR8213/CNRS -  ESPCI ParisTech - UPMC, 10 rue Vauquelin, 75005 Paris, France\\$^2$Institute for Nanoscale Physics and Chemistry, Katholieke Universiteit Leuven, Celestijnenlaan 200 D, B-3001 Heverlee, Belgium\\$^3$Laboratoire National des Champs Magn\'etiques Intenses, 143 Avenue de Rangueil, 31400 Toulouse, France\\$^4$National Physical Laboratory, Council of Scientific and Industrial ResearchNew Delhi 110012, and Indian Institute of Technology Kanpur, Kanpur 208016, India}

\begin{abstract}

Driving a two-dimensional superconductor normal by applying a high magnetic field may lead to Cooper pair localization. In this case, there should be a quantum critical point associated with specific scaling laws. Such a transition has been evidenced in a  number of low critical temperature superconducting thin films and has been suggested to occur also in high temperature cuprate superconductors.  Here we show experimental evidence for \textit{two }distinct quantum critical regimes (QCP) when applying perpendicular magnetic fields to underdoped \LSCO\ thin films. At intermediate values of the magnetic field ($\sim18T-20T$) , a "ghost" QCP is observed, for which the values of the related critical exponents point towards a fermionic -as opposed to bosonic- scenario.  At higher ($\sim37 T$) magnetic field, another QCP is observed, which suggests the existence of either a 2D/3D or a clean/dirty temperature crossover. 

\end{abstract}

\maketitle

\section{Article}

\indent

When a superconductor is driven to an insulating state by applying a magnetic field or increasing disorder, two different scenarios are possible. Conventionally, the Cooper pairs are first destroyed and then localized, which corresponds to a "fermionic" mechanism. But Ma and Lee \cite{Ma:1985te} were the first to suggest that the Cooper pairs might instead directly localize, leading to a "boson" localization mechanism. 

For the latter scenario, one of the models which is proposed for 2D systems is based on the duality between vortices and Cooper pairs and leads to the formation of a Bose insulator for a resistance per square exactly equal to the quantum of resistance  $R_C=h/4e^2$ \cite{Fisher:1990zz,Fisher:1990zza}, which corresponds to the existence of a quantum critical point (QCP) associated to a correlation length exponent $\nu>1$. The underlying microscopic mechanism may be the competition between Josephson coupling and Coulomb repulsion between superconducting domains, leading to localization of pairs \cite{Efetov:1980tr,Fazio:1991tw}.
For the "fermionic" scenario, a mechanism studied by Finkelstein and collaborators \cite{FinkelStein:1994ug} is that enhanced Coulomb repulsion decreases pairing and therefore the critical temperature $T_C$. Finally a third alternative has been proposed by Feigelman and coworkers  \cite{Feigelman:2001bw} and further developed by Spivak and coworkers  \cite{Spivak:2008cc} and applies to systems where the normal state is a dirty normal \textit{conductor}.
\indent
Since high-T$_C$ cuprate superconductors are known to be intrinsically lamellar, it has been proposed \cite{Steiner:2005} that a \textit{bosonic} superconductor/insulator transition is at play in such systems.

Here we show, after systematic investigation of the resistance versus temperature under high pulsed magnetic fields in a large variety of \LSCO\  thin films of different Sr concentrations, that the superconducting transition in an underdoped high T$_C$ superconductor reveals the existence of\textit{ two }quantum critical points.

\indent
The fluctuations associated to the first QCP are observable in a limited range of temperature and give way at low temperature to a second QCP with much higher critical magnetic field. The critical exponents associated to this first "ghost" QCP yields a value of $\nu$ not compatible with Boson localization theory, and $H_C$ is found to scale with $T_C$ with $1T=1K$. We propose two different scenarios to explain this double criticality. The first scenario is based on a 2D/3D temperature crossover. The second scenario is based on the proposal by Feigelman and Spivak  \cite{Feigelman:2001bw} \cite{Spivak:2008cc} of superconducting puddles coupled to each others through Josephson weak links, and questions the truly insulating character of the normal state of these systems. 

\indent
Surprisingly enough, such multiple criticality has recently been observed in 2D electron gas systems at the surface of $SrTiO_3$ \cite{Biscaras:2012vo}. 
We anticipate that this finding will stimulate further interest in the study of normal/superconducting phase transitions in 2D systems and will help clarifying the intrinsic versus extrinsic character of inhomogeneities in those systems.

\indent

Evidence has been found of boson localization in granular films of In/InOx by varying the composition \cite{Hebard:1984uj} or the magnetic field \cite{Hebard:1990up}, but also in homogeneous ultrathin films of Bi and Pb  \cite{Haviland:1989wi} by varying the layer thickness (which is equivalent as varying the disorder) and in MoGe \cite{Yazdani:1995ut}. On the contrary, there are evidences of absence of boson localization in amorphous InOx films \cite{Gantmakher:2000tp}. In a number of experiments, the findings are consistent with the boson localization approach but the critical exponents are close to the one expected for a classical (2+1)D xy model in the absence of disorder, for example in Bi \cite{Markovic:1999vy} and  NbSi \cite{Aubin:2006ju}.  In disordered amorphous thin films, a metallic state is observed just above the superconducting transition \cite{Mason:2001ed}. For a classification of the experimental results with normal state conductivity see\cite{Steiner:2008ke}.

As far as high T$_C$ cuprates are concerned, the disorder-tuned transition had originally been studied by Wang and coworkers \cite{Wang:1991vj} and under magnetic field in YBa$_2$Cu$_3$O$_{6.38}$ by Seidler et al. \cite{Seidler:1992ww}. Recently  Bollinger and coworkers \cite{Bollinger:2012eq} have reported SIT driven by an\textit{ electrical field} in  \LSCO monolayers or bilayers.   
Another group has reported similar measurements on \YBCO   thin films for a carrier concentration  $x\simeq0.05$ \cite{Leng:2011}.

We have investigated the behavior of resistance versus temperature under high magnetic fields in a large variety of typically 100nm-thick  \LSCO\  thin films of different Sr concentrations.  Part of the transport measurements were carried out at the KU Leuven high field facility and the other part at the Toulouse LNCMI high field facility.
Pulsed high-field measurements up to 49-58~T were performed from 1.5 or 4.2~K to 300~K on nine \LSCO thin films with different Sr concentration x. The c-axis oriented films were mounted in general with $\mu_0H // c$ (perpendicular field),  and the current (of typically mA and 50 kHz) was along the ab-plane (I // ab).  The reported data were obtained on these epitaxial films (of thickness t $\simeq$100  nm),  patterned in strips of 1000 x 500 $\mu$m, using four probe measurements.

\begin{figure}
\includegraphics[width=0.49\textwidth]{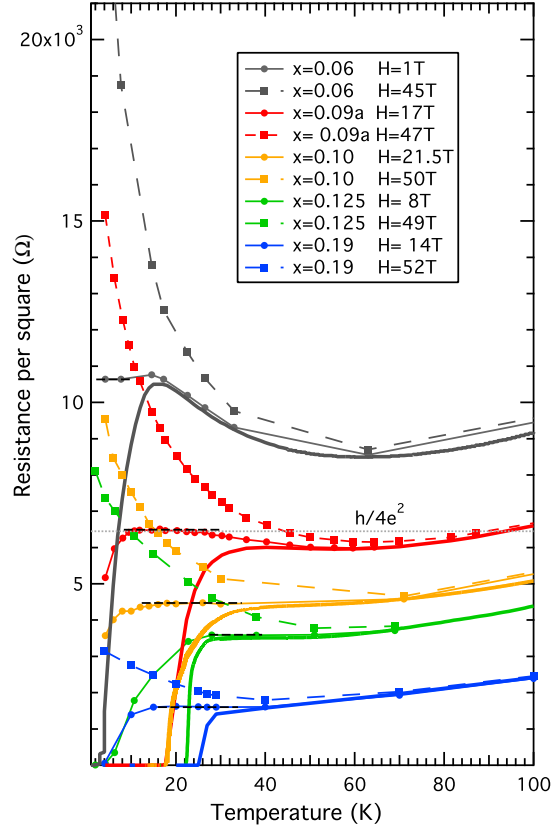}
\caption{Resistance per square of CuO$_2$ layer versus temperature for five representative samples. Solid lines H=0T; circles and lines:  H=H$_C$ (the values of H$_C$ are indicated in the legend); squares and dashed lines:  square resistance  at maximal field, as indicated in the legend. The plateaus are highlighted with dashed thick lines. The thin grey dotted line denotes the position of the quantum of resistance $h/4e^2$.}
\label{plat1}
\end{figure}

At maximal field (between 49~T and 57~T according to the different experiments),  the low temperature resistance of the underdoped samples versus temperature varies as $Ln(1/T)$ as was reported previously \cite{Ando:1995zz,Leridon:2007eb}. This is consistent with the conductivity of a Bose metal predicted by \cite{Das:1998tk}.
 It is noteworthy that for the underdoped samples that are \textit{not} superconducting at zero magnetic field (i.e. $x<0.06$), the resistance versus temperature at zero magnetic field exhibits a Shklovskii-Efros localization law ($\rho=\rho_0 exp((T/T_0)^{1/2}$) instead of a $Ln(1/T)$. (See the data in Weckuysen et al. \cite{Weckuysen:2002b} ).
This may be an indication that $\rho=\rho_0 Ln(1/T)$, whatever its physical origin,  probably related to granular metallicity,  is the strongest "insulating" behavior allowing superconductivity to develop \footnote{M. Grilli. Private communication}. 
In any case, in absence of a theory for the $Ln(T)$ behavior of the "normal state", it is questionable to assert that the normal state is insulating even though $dR/dT<0$.  Actually the normal state resistance of our samples is found to be of the order of $h/4e^2$ at the transition point (See Figure \ref{plat1}).

On overdoped samples ($x>0.2$) the R(H) curves measured in perpendicular magnetic field show no crossing point since no upturn of the resistance versus temperature is present down to 4K for fields as high as 50~T.  For this range of doping levels the system is seen metallic in our range of measurements.
We shall now focus on the underdoped and optimally doped samples.

In previous work from Ando and coworkers \cite{Ando:1995zz}, a negative magnetoresistance (MR) is shown at high magnetic field. This was argued to plead in favor of the presence of a Bose insulator by \cite{Steiner:2005}.  We did not observe such behaviour here in samples with similar Sr concentrations and in the same field range \footnote{Although the measured voltage did exhibit such a negative slope, when calculating the resistance by dividing by the effective current through the sample that was measured during the pulse, the negative MR was not present.}.  However, by contrast to the overdoped case, on every underdoped and optimally doped sample, at least one fixed crossing point in the $R(H)$ data was observed \textit{within a finite range of temperatures} for $H=H_C(x)$. (This crossing point was seen in a range of temperatures where the data in ref \cite{Ando:1995zz} are not shown.) This leads to a plateau in the R versus T curves when the magnetic field is equal to $H_C$.

\begin{figure}
\includegraphics[width=1\textwidth]{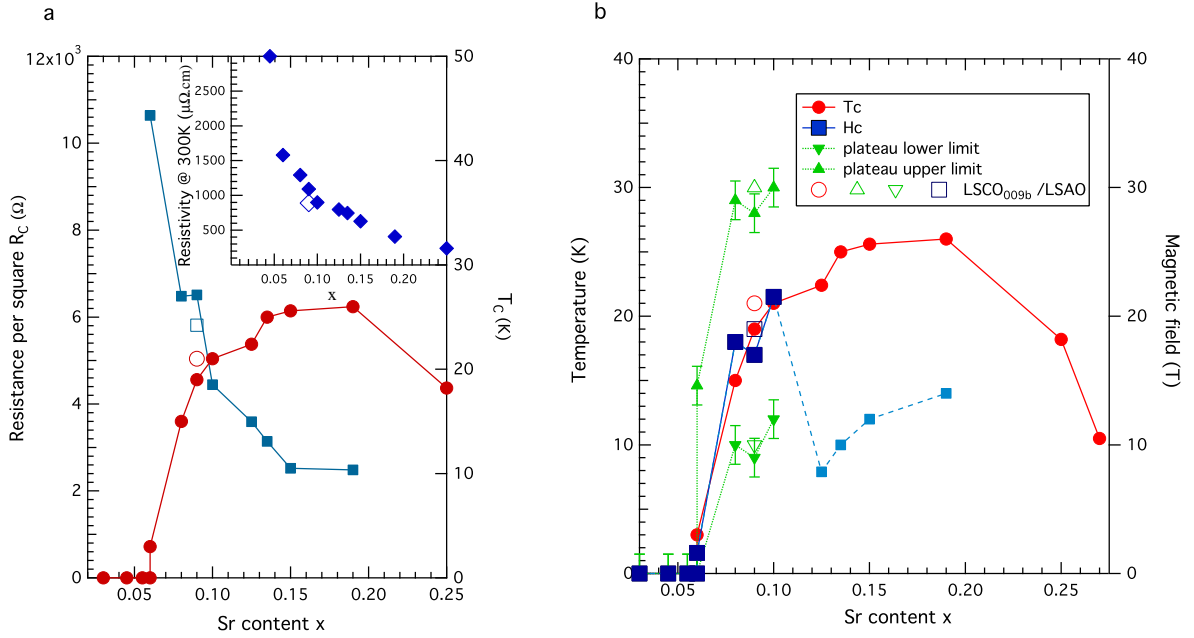}
\caption{\textbf{a}  Critical temperature $T_C$ (red dots, right scale) and square resistance $R_C$ (blue squares, left scale) at the plateau as a function of the Sr content. Inset: Room temperature resistivity ($\mu \Omega$.cm) as a function of Sr content x for all samples. Nota: open symbols refer to sample LSCO$_{009b}$ grown on LSAO.\textbf{b} Critical temperature $T_C$ (red dots, left scale) and critical field $H_C$ (blue squares, right scale), as a function of the Sr content $x$. Note that the scales for $H_C$ and $T_C$ are the same ($1T\approx1K$).  The down and up  triangles denote the respective lower and upper temperature limits for the observation of the plateaus, defined for a spread in the resistance values of less than 1\%. The variation of $H_C$ displays an abrupt change at about $x=0.125$, pointing toward different origins for the plateaus below and above $x=0.125$.The smaller squares indicate a different origin (residual resistivity) for the plateaus for $x\geq0.125$ since no scaling is associated (see text).}
\label{phadiag}
\end{figure}

In Figure \ref{plat1} are displayed the resistances per square of CuO$_2$ plane  versus temperature  for a selected set of five representative samples with x ranging from 0.06 to 0.19, for the field values $H=H_C$ (circles), $H=0~T$ (solid lines) and H$_{max}$ (squares).  The resistance at the plateau is depicted in Figure \ref{phadiag} \textbf{a}  as a function of the Sr content and varies monotonically. (The two samples LSCO$_{0.009a}$/STO and  LSCO$_{0.009b}$/LSAO grown on different substrates show some small discrepancies.)

Figure \ref{phadiag}\textbf{b} summarizes the values of $T_C$ taken at the transition inflection point (reds dots), $H_C$ (blue squares) and the temperature range of observation of the plateaus for different LSCO thin films as a function of the Sr content x.   The temperature range of observation of the plateau, denoted by up and down triangles corresponds to a spread in the resistance values of less than $1\%$. The error bars on the triangles correspond to the typical temperature spacing between different measurements. 

As may be seen from both Figures \ref{plat1} and \ref{phadiag} \textbf{b}, there is a great discrepancy between the data for $x<0.125$ and $x\geq0.125$.  $H_C$ increases with doping and scales with $T_C$ with 1K=1T for $x <0.125$,  then decreases abruptly at $x=0.125$ and then increases again.  Although for all  $x<0.125$ the plateau develops at temperatures well below the superconductivity onset temperature at zero field, on the contrary for  $x\geq0.125$ it is only present at temperatures above the foot of the transition at H=0T. 
Besides, for $x<0.125$ the resistance at the plateau is higher than the minimum of the full field resistivity curve and for $x\geq0.125$ it is slightly lower. Moreover, for samples with x$\geq0.125$, our attempt to obtain scaling of the data around the apparent crossing point - as will be described in the following paragraph, revealed to be unsuccessful.
These four observations point towards different explanations for the existence of the plateaus for x lower or greater than $1/8$.  Above this value, the apparent plateau is more likely to be related to the residual resistivity of the film. For the following, we will focus on the plateau observable for $x<0.125$.

\begin{figure}
\includegraphics[width=1\textwidth]{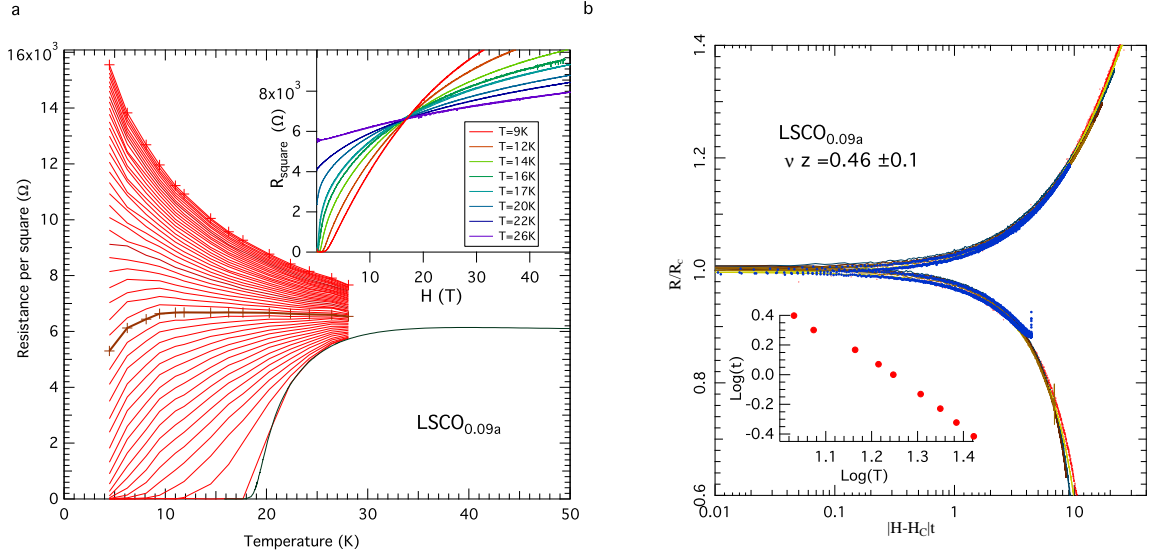}
\caption{\textbf{a}  R(T) data for different magnetic field values, ranging from 0~T to 47~T by steps of 1~T for sample LSCO$_{0.09a}$/STO ($x=0.09$). The brown line, showing a plateau from about 9~K to about 26~K, corresponds to 17~T.  Inset: Corresponding R(H) data for temperatures between 9K (lower curve at low fields) and 26 K (upper curve at low fields). \textbf{b}  Scaling of the R(H) curves for the same sample, for $R/R_C=f(|H-H_C|t)$ , with $t=T^{-1/\nu z}$ and $\nu z=0.46\pm0.1$.  Inset: Log(t) versus Log(T).}
\label{LS644}
\end{figure}

The inset of Figure \ref{LS644}a shows typical R(H) data curves for one of the samples whose Sr content is x=0.09 (LSCO$_{0.09a}$); a fixed point is observed for temperatures between  9~K and  26~K for $R_C=6.6 k\Omega,H_f=17~T$. The resistance per square of CuO plane versus temperature values are plotted in Fig. \ref{LS644}a for sample LSCO${0.09a}$. A plateau corresponding to the fixed point is clearly visible for H=17~T, for a square resistance of about $h/4e^2$. However a departure from the plateau at low temperature is also visible. This figure is typical of what was observed in all the samples with x ranging from 0.06 to 0.1.
The fact that the square resistance at the plateau  R$_C$ is almost equal to $h/4e^2$ in sample LSCO${0.09a}$ is rather fortuitous since R$_C$ is indeed found to vary with doping, as may be seen in Fig. \ref{plat1}. 

Scaling analysis was performed at the vicinity of the fixed point. Figure \ref{LS644}b shows the scaling of the same curves for sample LSCO$_{0.009a}$, as $R/{R_C}=f(|H-H_C|T^{-1/\nu z})$ with $\nu z = 0.46\pm0.1$.  The scaling exponent $\nu z$  is found to vary from 0.45$\pm 0.1$  to $0.63\pm0.1$ for the four different samples on which scaling was possible (x=0.08,0.09a, 0.09b and 0.1).  It is therefore possible to infer the values of $\nu$, assuming  z=1. They are much lower than the exponent predicted in the framework of a dirty boson picture ($\nu >1$), or  for classical percolation ($\nu=4/3$)  and than the exponent observed for the electrical field driven transition \cite{Bollinger:2012eq}  ($\nu z =3/2$). These values are lower - but still in rough agreement, with the exponent predicted for a (2D+1) quantum xy model ($\nu$z=2/3) and found for example in amorphous Bi \cite{Parendo:2005}, or $NbSi$ compounds \cite{Aubin:2006ju}. In any case these exponents are not compatible with a bosonic mechanism for the transition for which $\nu>1$.

As a further test, we performed the same measurements on sample LSCO$_{0.09a}$/STO ($x=0.09$) with the magnetic field  applied along the planes, and by contrast to the perpendicular case, the $R(H)$ curves do no longer show a crossing point in our range of measurement, which indicates that the transition is governed by a different mechanism in this case, and that the vortices are indeed relevant in the perpendicular case.

\begin{figure}
\includegraphics[width=1\textwidth]{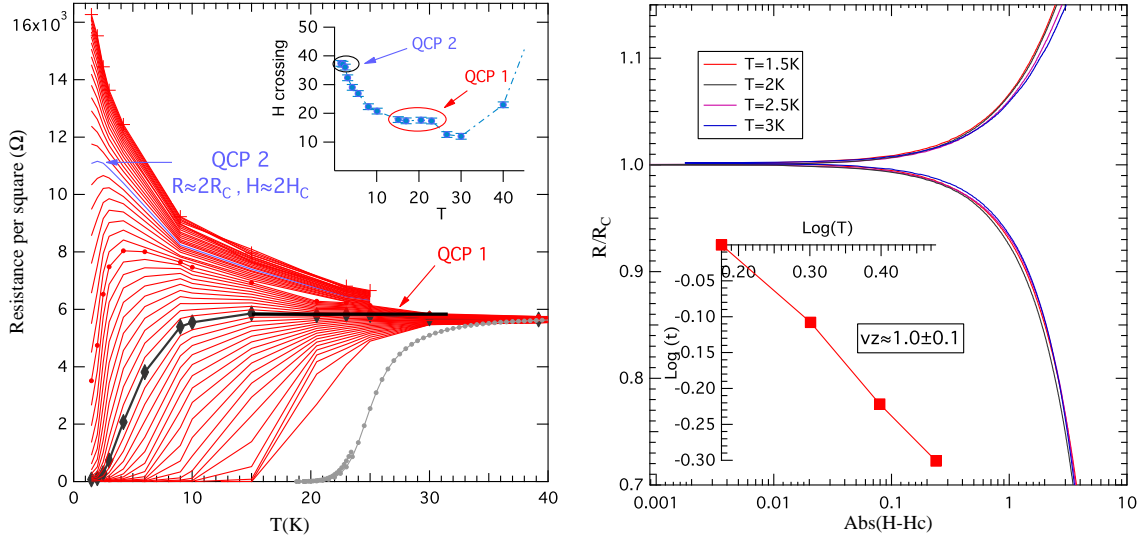}
\caption{\textbf{a} R(T) data for sample LSCO0.09b / LSAO (x=0.09) for different magnetic field values, ranging from 0~T to 56~T by steps of 1~T. The dark grey curve, exhibiting a plateau (marked in black) from less than 15~K up to about 30~K, corresponds to $H_C$=19~T.  The ranges of observation of the two QCP are designated by the arrows. (The resistance for magnetic field higher than 29~T (red circles)  was only measured for a reduced set of temperatures.) Inset: Value of the magnetic field corresponding to the intersection of two R(H) measurements taken at two consecutive temperatures as a function of the lowest temperature.
\textbf{b} Scaling of the low temperature data for the same sample, $R/R_C=f(|H-H_C|t)$ , with $t=T^{-1/\nu z}$ and $\nu z=1.0\pm0.1$. Inset: Log(t) versus Log(T).}
\label{LSCO009e}
\end{figure}

In order to better explore the low temperature behaviour,  we measured sample LSCO$_{0.09b}$/ LSAO $x=0.09$ (grown by PLD at IIT Kanpur on LSAO substrate) in the LNCMI high field facility from room temperature down to 1.5~K.   
(Due to compressive strain, samples grown on LSAO substrates are known to have  higher critical temperatures and lower normal state resistivities as compared to films grown on STO substrates, which is consistent with our findings.)  The above-described plateau was observed for about 19~T and the corresponding square resistance was found to be around 5.8 k$\Omega $.  The measurements carried out down to 1.5~K  at H$_C\simeq$19~T show that the system is indeed superconducting at T$\leq2$~K (See figure \ref{LSCO009e} a). These measurements also indicate that a low temperature insulator/superconductor transition takes place for $H\simeq 37~T$ (about 2$H_C$) and R~$\simeq 11000~ \Omega$ (about 2$R_C$) at low temperature.
The inset of Figure \ref{LSCO009e} a depicts the evolution of the magnetic field corresponding to the crossing point of two consecutive R(H) measurements as a function of the lowest of the two temperatures. It clearly shows the occurrence of two accumulation points at high and low temperature respectively. 
The second QCP is associated to a scaling law yielding critical exponents of about $\nu z\simeq 1.0\pm0.1$, i.e. about twice the exponents observed for QCP1, as can be seen in Figure \ref{LS644} b, therefore not discarding a bosonic scenario.

In most of the data reported in the literature, the resistance plateau is observed down to the lowest measurable temperature, presumably existing at zero temperature.  Our measurements evidence on the contrary the presence of two distinct QCP each of them being characterized by different critical exponents\footnote{Strictly speaking, QCP1 \textit{is not a real quantum critical point} but a "ghost" QCP since different physics takes over at low temperature.}.
We suggest two different explanations for this observation.

1) In contradiction to the monolayer studied by Bollinger and coworkers, our systems are not instrinsically 2D but rather \textit{lamellar} and this raises the possibility of a 2D-3D crossover.
As long as the c-axis  coherence length is smaller than the interlayer spacing (between the CuO$_2$ planes) then the system behaves like a 2D system in the Lawrence-Doniach sense \cite{Lawrence:1971}, but in a quantum way.  It may then be governed by a "ghost" 2D quantum critical point (QCP1) with critical exponents of a classical (2+1)D model. However, when the temperature is decreased towards $T_C^{2D}=0$ while cooling down exactly at $H_C(x)$,  at some point the coherence length gets larger than the interlayer spacing,  the system recovers classical 3D behaviour and the 2D QCP becomes irrelevant.  Then the 3D transition takes place at much higher field (about 2H$_C$) and much higher square resistance (about 2R$_C$ for sample LSCO$_{0.09b}$) and a second QCP (QCP2) is observed whose exponents should correspond to a (3+1)d xy model. 

2) The second interpretation is based on the model proposed by Feigel'man \cite{Feigelman:2001bw} and Spivak  \cite{Spivak:2008cc} and has been recently applied to the 2D electron gas created at perovskite interfaces \cite{Biscaras:2012vo}.  In this model, superconducting puddles of size d are coupled to each others through Josephson weak links of lengths b, but the essential difference is that, due to quantum fluctuations,  the superconductivity inside the puddles is \textit{unstable} at $T=0$ for a critical value of the conductance of the normal conductor less than $g_c \simeq 0.1 ln^2(b/d)$. In our case, the crossover between the two QCP is governed by the dephasing length as defined by  \cite{Sondhi:1997wz} which varies as $L_\phi\sim T^{-1/z}$. At high temperature this length is smaller than the typical size d of a superconducting puddle. The puddles are therefore decoupled and the system inside a given puddle thus behaves like a (2+1)D xy model in the clean limit (according to Harris criteria \cite{Harris:1974tn}) with a critical exponent close to $\nu z=2/3$. The observed critical field scales with $T_C$.
At lower temperature, $L_\phi\gg d$ ; the puddles are coupled and the system is inhomogeneous. Remarkably enough, the overall critical field of this inhomogeneous system is predicted to be higher than the critical field of one superconducting puddle, which is consistent with our observations. The transition is then governed by a critical exponent $\nu\ge1$, corresponding to a dirty limit for the fluctuations \cite{Harris:1974tn}. 
This mechanism thus implies that the normal state in our systems is a dirty conductor rather than an insulator, which is consistent with the low temperature behaviour of the resistivity $R\sim Ln(T)$. Theory predicts that the resistance of a two-dimensional electron gas at zero temperature should be either zero or infinite \cite{Ma:1985te} \cite{Lee:1985zzc} excluding any intermediate metallic behavior.  
However, due to the finite volume of the samples, the effects of localization at $T>0$ are almost unobservable  \cite{Spivak:2008cc}, therefore even for a thin film (and even more so for a lamellar compound), this dirty conductor scenario should not be disregarded.

As a conclusion, we have investigated the magnetic field driven superconductive transition in eleven samples of \LSCO\ thin films with various Sr content x, under high pulsed magnetic field. Our results indicate the presence of a "ghost" QCP for every underdoped superconducting sample with $0.06 \leq x\leq0.1$, associated  with critical exponent $\nu z \simeq 0.5-0.6$, therefore not compatible with boson localization. The critical field is found to scale with the $T_C$ with $1T\simeq1K$. A second QCP is observed at lower temperature and higher magnetic field with $\nu z\simeq1$. Two possible explanations are proposed based either on the existence of a 2D/3D crossover due to the lamellar nature of the system or to a clean/dirty crossover with two QCP as described by \cite{Feigelman:2001bw,Spivak:2008cc}. In any case, all the observations are consistent with a\textit{ fermionic } mechanism for the transition at the "ghost" QCP and the presence of an extended dirty metallic state is not ruled out.
We think that this system constitutes an example of \textit{ electronically textured} material, the superconducting domains being controlled by a phase coherence length. Remarkably enough our findings are similar to recent observations in 2D electron gas at interfaces \cite{Biscaras:2012vo}.
We believe that our findings  will stimulate further interest in the field of inhomogeneous superconductivity and phase transitions associated with multiple criticality.

\section{Methods:}

The typically 100nm-thick La$_{2-x}$Sr$_x$CuO$_4$ films were prepared either  by DC magnetron sputtering on SrTiO$_4$ (STO) substrates at KU Leuven excepted one of them  (LSCO$_{0.09b}$) that was deposited by pulsed laser deposition at IIT Kanpur on LaSrAlO$_4$ (LSAO). 
The quality of the films was checked by XRD which showed good epitaxial growth, absence of impurity phases, with rocking curves exhibiting peaks of FWHM equal to about 0.3 degrees. 

The critical temperatures $T_C$ were determined  as the transition mid-point temperature obtained by measuring the resistance under zero magnetic field. The measured values are fully consistent with previously reported  measurements on thin films with same composition on STO \cite{Locquet:1996tt}. Optimal T$_C$ is usually found equal to 20K for samples of 40-60 nm and equal to 27 K for samples of 200 nm. Our samples are typically 100 nm thick with area $5\times5 mm^2$ and the optimal Tc is about 25K so consistent with the data shown in  Fig.4 in \cite{Locquet:1996tt}. For the sample grown at IIT Kanpur, the T$_C$ is somewhat higher, as expected for LSAO substrates and consistent with the findings in  \cite{Bozovic:2002da}. 

\section{Author contributions:}

JV, RP and RCB prepared the samples. BL, JV, VVM and BV performed the high-field measurements. BL interpreted the data and wrote the manuscript. All authors commented on, discussed and improved the manuscript.

\acknowledgments
The authors acknowledge for assistance during the experiments, G. Scheerer at LNCMI and T. Wambecq at KU Leuven, and for fruitful discussions, M. Grilli and  S. Caprara. The work at the KU Leuven has been suported by the FWO Programmes and Methusalem Funding by the Flemish Government.  Research at IIT Kanpur has been supported by the J.C. Bose National Fellowship (R.C. Budhani). Part of this work has been founded by EuroMagNET II under the EU contract number 228043 and has also been supported through a SESAME grant by the Ile-de-France Regional Counsel.

\section{Competing financial interests}
The authors declare no competing financial interests.


\begin{thebibliography}{34}
\expandafter\ifx\csname natexlab\endcsname\relax\def\natexlab#1{#1}\fi
\expandafter\ifx\csname bibnamefont\endcsname\relax
  \def\bibnamefont#1{#1}\fi
\expandafter\ifx\csname bibfnamefont\endcsname\relax
  \def\bibfnamefont#1{#1}\fi
\expandafter\ifx\csname citenamefont\endcsname\relax
  \def\citenamefont#1{#1}\fi
\expandafter\ifx\csname url\endcsname\relax
  \def\url#1{\texttt{#1}}\fi
\expandafter\ifx\csname urlprefix\endcsname\relax\def\urlprefix{URL }\fi
\providecommand{\bibinfo}[2]{#2}
\providecommand{\eprint}[2][]{\url{#2}}

\bibitem[{\citenamefont{Ma and Lee}(1985)}]{Ma:1985te}
\bibinfo{author}{\bibfnamefont{M.}~\bibnamefont{Ma}} \bibnamefont{and}
  \bibinfo{author}{\bibfnamefont{P.~A.} \bibnamefont{Lee}},
  \bibinfo{journal}{Physical Review B} \textbf{\bibinfo{volume}{32}},
  \bibinfo{pages}{5658} (\bibinfo{year}{1985}).

\bibitem[{\citenamefont{Fisher}(1990)}]{Fisher:1990zz}
\bibinfo{author}{\bibfnamefont{M.~P.~A.} \bibnamefont{Fisher}},
  \bibinfo{journal}{Physical Review Letters} \textbf{\bibinfo{volume}{65}},
  \bibinfo{pages}{923} (\bibinfo{year}{1990}).

\bibitem[{\citenamefont{Fisher et~al.}(1990)\citenamefont{Fisher, Grinstein,
  and Girvin}}]{Fisher:1990zza}
\bibinfo{author}{\bibfnamefont{M.~P.~A.} \bibnamefont{Fisher}},
  \bibinfo{author}{\bibfnamefont{G.}~\bibnamefont{Grinstein}},
  \bibnamefont{and} \bibinfo{author}{\bibfnamefont{S.~M.}
  \bibnamefont{Girvin}}, \bibinfo{journal}{Physical Review Letters}
  \textbf{\bibinfo{volume}{64}}, \bibinfo{pages}{587} (\bibinfo{year}{1990}).

\bibitem[{\citenamefont{Efetov}(1980)}]{Efetov:1980tr}
\bibinfo{author}{\bibfnamefont{K.~B.} \bibnamefont{Efetov}},
  \bibinfo{journal}{Sov. Phys.-JETP (Engl. Transl.);(United States)}
  \textbf{\bibinfo{volume}{51}} (\bibinfo{year}{1980}).

\bibitem[{\citenamefont{Fazio and Sch{\"o}n}(1991)}]{Fazio:1991tw}
\bibinfo{author}{\bibfnamefont{R.}~\bibnamefont{Fazio}} \bibnamefont{and}
  \bibinfo{author}{\bibfnamefont{G.}~\bibnamefont{Sch{\"o}n}},
  \bibinfo{journal}{Physical Review B} \textbf{\bibinfo{volume}{43}},
  \bibinfo{pages}{5307} (\bibinfo{year}{1991}).

\bibitem[{\citenamefont{Finkel'Stein}(1994)}]{FinkelStein:1994ug}
\bibinfo{author}{\bibfnamefont{A.~M.} \bibnamefont{Finkel'Stein}},
  \bibinfo{journal}{Physica B: Physics of Condensed Matter}
  \textbf{\bibinfo{volume}{197}}, \bibinfo{pages}{636} (\bibinfo{year}{1994}).

\bibitem[{\citenamefont{Feigel'man et~al.}(2001)\citenamefont{Feigel'man,
  Larkin, and Skvortsov}}]{Feigelman:2001bw}
\bibinfo{author}{\bibfnamefont{M.}~\bibnamefont{Feigel'man}},
  \bibinfo{author}{\bibfnamefont{A.}~\bibnamefont{Larkin}}, \bibnamefont{and}
  \bibinfo{author}{\bibfnamefont{M.}~\bibnamefont{Skvortsov}},
  \bibinfo{journal}{Physical Review Letters} \textbf{\bibinfo{volume}{86}},
  \bibinfo{pages}{1869} (\bibinfo{year}{2001}).

\bibitem[{\citenamefont{Spivak et~al.}(2008)\citenamefont{Spivak, Oreto, and
  Kivelson}}]{Spivak:2008cc}
\bibinfo{author}{\bibfnamefont{B.}~\bibnamefont{Spivak}},
  \bibinfo{author}{\bibfnamefont{P.}~\bibnamefont{Oreto}}, \bibnamefont{and}
  \bibinfo{author}{\bibfnamefont{S.~A.} \bibnamefont{Kivelson}},
  \bibinfo{journal}{Physical Review B} \textbf{\bibinfo{volume}{77}},
  \bibinfo{pages}{214523} (\bibinfo{year}{2008}).

\bibitem[{\citenamefont{Steiner et~al.}(2005)\citenamefont{Steiner, Boebinger,
  and Kapitulnik}}]{Steiner:2005}
\bibinfo{author}{\bibfnamefont{M.~A.} \bibnamefont{Steiner}},
  \bibinfo{author}{\bibfnamefont{G.}~\bibnamefont{Boebinger}},
  \bibnamefont{and}
  \bibinfo{author}{\bibfnamefont{A.}~\bibnamefont{Kapitulnik}},
  \bibinfo{journal}{Physical Review Letters} \textbf{\bibinfo{volume}{94}},
  \bibinfo{pages}{107008} (\bibinfo{year}{2005}).

\bibitem[{\citenamefont{Biscaras et~al.}(2012)\citenamefont{Biscaras, Bergeal,
  Hurand, Feuillet-Palma, Rastogi, Budhani, Grilli, Caprara, and
  Lesueur}}]{Biscaras:2012vo}
\bibinfo{author}{\bibfnamefont{J.}~\bibnamefont{Biscaras}},
  \bibinfo{author}{\bibfnamefont{N.}~\bibnamefont{Bergeal}},
  \bibinfo{author}{\bibfnamefont{S.}~\bibnamefont{Hurand}},
  \bibinfo{author}{\bibfnamefont{C.}~\bibnamefont{Feuillet-Palma}},
  \bibinfo{author}{\bibfnamefont{A.}~\bibnamefont{Rastogi}},
  \bibinfo{author}{\bibfnamefont{R.~C.} \bibnamefont{Budhani}},
  \bibinfo{author}{\bibfnamefont{M.}~\bibnamefont{Grilli}},
  \bibinfo{author}{\bibfnamefont{S.}~\bibnamefont{Caprara}}, \bibnamefont{and}
  \bibinfo{author}{\bibfnamefont{J.}~\bibnamefont{Lesueur}},
  \bibinfo{journal}{arXiv.org} \textbf{\bibinfo{volume}{cond-mat.supr-con}}
  (\bibinfo{year}{2012}).

\bibitem[{\citenamefont{Hebard and Paalanen}(1984)}]{Hebard:1984uj}
\bibinfo{author}{\bibfnamefont{A.~F.} \bibnamefont{Hebard}} \bibnamefont{and}
  \bibinfo{author}{\bibfnamefont{M.~A.} \bibnamefont{Paalanen}},
  \bibinfo{journal}{Physical Review B} \textbf{\bibinfo{volume}{30}},
  \bibinfo{pages}{4063} (\bibinfo{year}{1984}).

\bibitem[{\citenamefont{Hebard and Paalanen}(1990)}]{Hebard:1990up}
\bibinfo{author}{\bibfnamefont{A.~F.} \bibnamefont{Hebard}} \bibnamefont{and}
  \bibinfo{author}{\bibfnamefont{M.~A.} \bibnamefont{Paalanen}},
  \bibinfo{journal}{Physical Review Letters} \textbf{\bibinfo{volume}{65}},
  \bibinfo{pages}{927} (\bibinfo{year}{1990}).

\bibitem[{\citenamefont{Haviland et~al.}(1989)\citenamefont{Haviland, Liu, and
  Goldman}}]{Haviland:1989wi}
\bibinfo{author}{\bibfnamefont{D.~B.} \bibnamefont{Haviland}},
  \bibinfo{author}{\bibfnamefont{Y.}~\bibnamefont{Liu}}, \bibnamefont{and}
  \bibinfo{author}{\bibfnamefont{A.~M.} \bibnamefont{Goldman}},
  \bibinfo{journal}{Physical Review Letters} \textbf{\bibinfo{volume}{62}},
  \bibinfo{pages}{2180} (\bibinfo{year}{1989}).

\bibitem[{\citenamefont{Yazdani and Kapitulnik}(1995)}]{Yazdani:1995ut}
\bibinfo{author}{\bibfnamefont{A.}~\bibnamefont{Yazdani}} \bibnamefont{and}
  \bibinfo{author}{\bibfnamefont{A.}~\bibnamefont{Kapitulnik}},
  \bibinfo{journal}{Physical Review Letters} \textbf{\bibinfo{volume}{74}},
  \bibinfo{pages}{3037} (\bibinfo{year}{1995}).

\bibitem[{\citenamefont{Gantmakher et~al.}(2000)\citenamefont{Gantmakher,
  Golubkov, Dolgopolov, Shashkin, and Tsydynzhapov}}]{Gantmakher:2000tp}
\bibinfo{author}{\bibfnamefont{V.~F.} \bibnamefont{Gantmakher}},
  \bibinfo{author}{\bibfnamefont{M.~V.} \bibnamefont{Golubkov}},
  \bibinfo{author}{\bibfnamefont{V.~T.} \bibnamefont{Dolgopolov}},
  \bibinfo{author}{\bibfnamefont{A.}~\bibnamefont{Shashkin}}, \bibnamefont{and}
  \bibinfo{author}{\bibfnamefont{G.~E.} \bibnamefont{Tsydynzhapov}},
  \bibinfo{journal}{JETP Letters} \textbf{\bibinfo{volume}{71}},
  \bibinfo{pages}{473} (\bibinfo{year}{2000}).

\bibitem[{\citenamefont{Markovi{\'c} et~al.}(1999)\citenamefont{Markovi{\'c},
  Christiansen, Mack, Huber, and Goldman}}]{Markovic:1999vy}
\bibinfo{author}{\bibfnamefont{N.}~\bibnamefont{Markovi{\'c}}},
  \bibinfo{author}{\bibfnamefont{C.}~\bibnamefont{Christiansen}},
  \bibinfo{author}{\bibfnamefont{A.~M.} \bibnamefont{Mack}},
  \bibinfo{author}{\bibfnamefont{W.~H.} \bibnamefont{Huber}}, \bibnamefont{and}
  \bibinfo{author}{\bibfnamefont{A.~M.} \bibnamefont{Goldman}},
  \bibinfo{journal}{Physical Review B} \textbf{\bibinfo{volume}{60}},
  \bibinfo{pages}{4320} (\bibinfo{year}{1999}).

\bibitem[{\citenamefont{Aubin et~al.}(2006)\citenamefont{Aubin,
  Marrache-Kikuchi, Pourret, Behnia, Berg{\'e}, Dumoulin, and
  Lesueur}}]{Aubin:2006ju}
\bibinfo{author}{\bibfnamefont{H.}~\bibnamefont{Aubin}},
  \bibinfo{author}{\bibfnamefont{C.}~\bibnamefont{Marrache-Kikuchi}},
  \bibinfo{author}{\bibfnamefont{A.}~\bibnamefont{Pourret}},
  \bibinfo{author}{\bibfnamefont{K.}~\bibnamefont{Behnia}},
  \bibinfo{author}{\bibfnamefont{L.}~\bibnamefont{Berg{\'e}}},
  \bibinfo{author}{\bibfnamefont{L.}~\bibnamefont{Dumoulin}}, \bibnamefont{and}
  \bibinfo{author}{\bibfnamefont{J.}~\bibnamefont{Lesueur}},
  \bibinfo{journal}{Physical Review B} \textbf{\bibinfo{volume}{73}},
  \bibinfo{pages}{094521} (\bibinfo{year}{2006}).

\bibitem[{\citenamefont{Mason and Kapitulnik}(2001)}]{Mason:2001ed}
\bibinfo{author}{\bibfnamefont{N.}~\bibnamefont{Mason}} \bibnamefont{and}
  \bibinfo{author}{\bibfnamefont{A.}~\bibnamefont{Kapitulnik}},
  \bibinfo{journal}{Physical Review B} \textbf{\bibinfo{volume}{64}},
  \bibinfo{pages}{060504} (\bibinfo{year}{2001}).

\bibitem[{\citenamefont{Steiner et~al.}(2008)\citenamefont{Steiner, Breznay,
  and Kapitulnik}}]{Steiner:2008ke}
\bibinfo{author}{\bibfnamefont{M.}~\bibnamefont{Steiner}},
  \bibinfo{author}{\bibfnamefont{N.}~\bibnamefont{Breznay}}, \bibnamefont{and}
  \bibinfo{author}{\bibfnamefont{A.}~\bibnamefont{Kapitulnik}},
  \bibinfo{journal}{Physical Review B} \textbf{\bibinfo{volume}{77}},
  \bibinfo{pages}{212501} (\bibinfo{year}{2008}).

\bibitem[{\citenamefont{Wang et~al.}(1991)\citenamefont{Wang, Beauchamp,
  Berkley, Johnson, Liu, Zhang, and Goldman}}]{Wang:1991vj}
\bibinfo{author}{\bibfnamefont{T.}~\bibnamefont{Wang}},
  \bibinfo{author}{\bibfnamefont{K.~M.} \bibnamefont{Beauchamp}},
  \bibinfo{author}{\bibfnamefont{D.~D.} \bibnamefont{Berkley}},
  \bibinfo{author}{\bibfnamefont{B.~R.} \bibnamefont{Johnson}},
  \bibinfo{author}{\bibfnamefont{J.~X.} \bibnamefont{Liu}},
  \bibinfo{author}{\bibfnamefont{J.}~\bibnamefont{Zhang}}, \bibnamefont{and}
  \bibinfo{author}{\bibfnamefont{A.~M.} \bibnamefont{Goldman}},
  \bibinfo{journal}{Physical Review B} \textbf{\bibinfo{volume}{43}},
  \bibinfo{pages}{8623} (\bibinfo{year}{1991}).

\bibitem[{\citenamefont{Seidler et~al.}(1992)\citenamefont{Seidler, Rosenbaum,
  and Veal}}]{Seidler:1992ww}
\bibinfo{author}{\bibfnamefont{G.~T.} \bibnamefont{Seidler}},
  \bibinfo{author}{\bibfnamefont{T.~F.} \bibnamefont{Rosenbaum}},
  \bibnamefont{and} \bibinfo{author}{\bibfnamefont{B.~W.} \bibnamefont{Veal}},
  \bibinfo{journal}{Physical Review, B: Condensed Matter;(United States)}
  \textbf{\bibinfo{volume}{45}} (\bibinfo{year}{1992}).

\bibitem[{\citenamefont{Bollinger et~al.}(2012)\citenamefont{Bollinger, Dubuis,
  Yoon, Pavuna, Misewich, and Bozˇovic}}]{Bollinger:2012eq}
\bibinfo{author}{\bibfnamefont{A.~T.} \bibnamefont{Bollinger}},
  \bibinfo{author}{\bibfnamefont{G.}~\bibnamefont{Dubuis}},
  \bibinfo{author}{\bibfnamefont{J.}~\bibnamefont{Yoon}},
  \bibinfo{author}{\bibfnamefont{D.}~\bibnamefont{Pavuna}},
  \bibinfo{author}{\bibfnamefont{J.}~\bibnamefont{Misewich}}, \bibnamefont{and}
  \bibinfo{author}{\bibfnamefont{I.}~\bibnamefont{Bozˇovic}},
  \bibinfo{journal}{Nature} \textbf{\bibinfo{volume}{472}},
  \bibinfo{pages}{458} (\bibinfo{year}{2012}).

\bibitem[{\citenamefont{Leng et~al.}(2011)\citenamefont{Leng,
  Garcia-Barriocanal, Bose, Lee, and Goldman}}]{Leng:2011}
\bibinfo{author}{\bibfnamefont{X.}~\bibnamefont{Leng}},
  \bibinfo{author}{\bibfnamefont{J.}~\bibnamefont{Garcia-Barriocanal}},
  \bibinfo{author}{\bibfnamefont{S.}~\bibnamefont{Bose}},
  \bibinfo{author}{\bibfnamefont{Y.}~\bibnamefont{Lee}}, \bibnamefont{and}
  \bibinfo{author}{\bibfnamefont{A.~M.} \bibnamefont{Goldman}},
  \bibinfo{journal}{Physical Review Letters} \textbf{\bibinfo{volume}{107}},
  \bibinfo{pages}{027001} (\bibinfo{year}{2011}).

\bibitem[{\citenamefont{Ando et~al.}(1995)\citenamefont{Ando, Boebinger,
  Passner, Kimura, and Kishio}}]{Ando:1995zz}
\bibinfo{author}{\bibfnamefont{Y.}~\bibnamefont{Ando}},
  \bibinfo{author}{\bibfnamefont{G.~S.} \bibnamefont{Boebinger}},
  \bibinfo{author}{\bibfnamefont{A.}~\bibnamefont{Passner}},
  \bibinfo{author}{\bibfnamefont{T.}~\bibnamefont{Kimura}}, \bibnamefont{and}
  \bibinfo{author}{\bibfnamefont{K.}~\bibnamefont{Kishio}},
  \bibinfo{journal}{Physical Review Letters} \textbf{\bibinfo{volume}{75}},
  \bibinfo{pages}{4662} (\bibinfo{year}{1995}).

\bibitem[{\citenamefont{Leridon et~al.}(2007)\citenamefont{Leridon, Vanacken,
  Wambecq, and Moshchalkov}}]{Leridon:2007eb}
\bibinfo{author}{\bibfnamefont{B.}~\bibnamefont{Leridon}},
  \bibinfo{author}{\bibfnamefont{J.}~\bibnamefont{Vanacken}},
  \bibinfo{author}{\bibfnamefont{T.}~\bibnamefont{Wambecq}}, \bibnamefont{and}
  \bibinfo{author}{\bibfnamefont{V.}~\bibnamefont{Moshchalkov}},
  \bibinfo{journal}{Physical Review B} \textbf{\bibinfo{volume}{76}},
  \bibinfo{pages}{012503} (\bibinfo{year}{2007}).

\bibitem[{\citenamefont{Das and Doniach}(1998)}]{Das:1998tk}
\bibinfo{author}{\bibfnamefont{D.}~\bibnamefont{Das}} \bibnamefont{and}
  \bibinfo{author}{\bibfnamefont{S.}~\bibnamefont{Doniach}},
  \bibinfo{journal}{Physical Review B} \textbf{\bibinfo{volume}{57}},
  \bibinfo{pages}{14440} (\bibinfo{year}{1998}).

\bibitem[{\citenamefont{Weckhuysen}(2002)}]{Weckuysen:2002b}
\bibinfo{author}{\bibfnamefont{L.}~\bibnamefont{Weckhuysen}},
  \bibinfo{journal}{{PhD thesis: High magnetic field study of the transport
  properties of underdoped and overdoped LSCO epitaxial thin films}}
  (\bibinfo{year}{2002}).

\bibitem[{\citenamefont{Parendo et~al.}(2005)\citenamefont{Parendo, Tan,
  Bhattacharya, Eblen-Zayas, Staley, and Goldman}}]{Parendo:2005}
\bibinfo{author}{\bibfnamefont{K.~A.} \bibnamefont{Parendo}},
  \bibinfo{author}{\bibfnamefont{K.~H. S.~B.} \bibnamefont{Tan}},
  \bibinfo{author}{\bibfnamefont{A.}~\bibnamefont{Bhattacharya}},
  \bibinfo{author}{\bibfnamefont{M.}~\bibnamefont{Eblen-Zayas}},
  \bibinfo{author}{\bibfnamefont{N.~E.} \bibnamefont{Staley}},
  \bibnamefont{and} \bibinfo{author}{\bibfnamefont{A.~M.}
  \bibnamefont{Goldman}}, \bibinfo{journal}{Physical Review Letters}
  \textbf{\bibinfo{volume}{94}}, \bibinfo{pages}{197004}
  (\bibinfo{year}{2005}).

\bibitem[{\citenamefont{Lawrence and Doniach}(1971)}]{Lawrence:1971}
\bibinfo{author}{\bibfnamefont{W.}~\bibnamefont{Lawrence}} \bibnamefont{and}
  \bibinfo{author}{\bibfnamefont{S.}~\bibnamefont{Doniach}},
  \bibinfo{journal}{Proc. 12th Int. Conf. on Low Temp. Phys.} p.
  \bibinfo{pages}{361} (\bibinfo{year}{1971}).

\bibitem[{\citenamefont{Sondhi et~al.}(1997)\citenamefont{Sondhi, Girvin,
  Carini, and Shahar}}]{Sondhi:1997wz}
\bibinfo{author}{\bibfnamefont{S.~L.} \bibnamefont{Sondhi}},
  \bibinfo{author}{\bibfnamefont{S.~M.} \bibnamefont{Girvin}},
  \bibinfo{author}{\bibfnamefont{J.~P.} \bibnamefont{Carini}},
  \bibnamefont{and} \bibinfo{author}{\bibfnamefont{D.}~\bibnamefont{Shahar}},
  \bibinfo{journal}{Reviews of Modern Physics} \textbf{\bibinfo{volume}{69}},
  \bibinfo{pages}{315} (\bibinfo{year}{1997}).

\bibitem[{\citenamefont{Harris}(1974)}]{Harris:1974tn}
\bibinfo{author}{\bibfnamefont{A.~B.} \bibnamefont{Harris}},
  \bibinfo{journal}{J.Phys.} \textbf{\bibinfo{volume}{7}},
  \bibinfo{pages}{1671} (\bibinfo{year}{1974}).

\bibitem[{\citenamefont{Lee and Ramakrishnan}(1985)}]{Lee:1985zzc}
\bibinfo{author}{\bibfnamefont{P.~A.} \bibnamefont{Lee}} \bibnamefont{and}
  \bibinfo{author}{\bibfnamefont{T.~V.} \bibnamefont{Ramakrishnan}},
  \bibinfo{journal}{Reviews of Modern Physics} \textbf{\bibinfo{volume}{57}},
  \bibinfo{pages}{287} (\bibinfo{year}{1985}).

\bibitem[{\citenamefont{Locquet et~al.}(1996)\citenamefont{Locquet, Jaccard,
  Cretton, Williams, Arrouy, M{\"a}chler, Schneider, and
  Martinoli}}]{Locquet:1996tt}
\bibinfo{author}{\bibfnamefont{J.-P.} \bibnamefont{Locquet}},
  \bibinfo{author}{\bibfnamefont{Y.}~\bibnamefont{Jaccard}},
  \bibinfo{author}{\bibfnamefont{A.}~\bibnamefont{Cretton}},
  \bibinfo{author}{\bibfnamefont{E.~J.} \bibnamefont{Williams}},
  \bibinfo{author}{\bibfnamefont{F.}~\bibnamefont{Arrouy}},
  \bibinfo{author}{\bibfnamefont{E.}~\bibnamefont{M{\"a}chler}},
  \bibinfo{author}{\bibfnamefont{T.}~\bibnamefont{Schneider}},
  \bibnamefont{and}
  \bibinfo{author}{\bibfnamefont{P.}~\bibnamefont{Martinoli}},
  \bibinfo{journal}{Physical Review B} \textbf{\bibinfo{volume}{54}},
  \bibinfo{pages}{7481} (\bibinfo{year}{1996}).

\bibitem[{\citenamefont{Bozovic et~al.}(2002)\citenamefont{Bozovic, Logvenov,
  Belca, Narimbetov, and Sveklo}}]{Bozovic:2002da}
\bibinfo{author}{\bibfnamefont{I.}~\bibnamefont{Bozovic}},
  \bibinfo{author}{\bibfnamefont{G.}~\bibnamefont{Logvenov}},
  \bibinfo{author}{\bibfnamefont{I.}~\bibnamefont{Belca}},
  \bibinfo{author}{\bibfnamefont{B.}~\bibnamefont{Narimbetov}},
  \bibnamefont{and} \bibinfo{author}{\bibfnamefont{I.}~\bibnamefont{Sveklo}},
  \bibinfo{journal}{Physical Review Letters} \textbf{\bibinfo{volume}{89}},
  \bibinfo{pages}{107001} (\bibinfo{year}{2002}).

\end{thebibliography}

\end{document}